\documentclass{aa}
\usepackage{times,epsfig,amssymb,amsmath}

\newcommand{\chandra}{{\it Chandra}}
\newcommand{\xmm}{{\it XMM-Newton}}

\title{The outer regions of galaxy clusters: \\
\chandra\ constraints on the X-ray surface brightness}
\titlerunning{The outer regions of galaxy clusters}

\author{S. Ettori\inst{1,2} \and I. Balestra\inst{3}
} 
\authorrunning{S. Ettori et al.}

\institute{
 INAF, Osservatorio Astronomico di Bologna, via Ranzani 1, I-40127 Bologna, Italy
 \and INFN, Sezione di Bologna, viale Berti Pichat 6/2, I-40127 Bologna, Italy
 \and Max-Planck-Institut f\"ur extraterrestrische Physik, Giessenbachstr. 1, D-85748 Garching, Germany
}

\offprints{S. Ettori}

\mail{stefano.ettori@oabo.inaf.it}

\date{In press}

\begin{document}

\abstract
{We study the properties of the X-ray surface brightness profiles in a sample of galaxy 
clusters that were observed with \chandra\ and have emission detectable with a signal-to-noise
ratio higher than 2 per radial bin at a radius beyond $R_{500} \approx 0.7 \times R_{200}$.
}
{Our study aims to measure the slopes in both the X-ray surface brightness and gas density
profiles in the outskirts of massive clusters. These constraints are compared
with similar results obtained from observations and numerical simulations of the temperature
and dark-matter density profiles with the intention of presenting a consistent
picture of the outer regions of galaxy clusters.
 } 
{We extract the surface brightness profiles $S_b(r)$ of 52 X-ray luminous galaxy clusters at $z>0.3$
from X-ray exposures obtained with \chandra. 
These objects, which are of both high X-ray surface brightness and high redshift,
allow us to use \chandra\ either in ACIS-I or even ACIS-S configuration
to survey the cluster outskirts.
We estimate $R_{200}$ using both a $\beta-$model that reproduces
the surface brightness profiles and scaling relations from the literature.
The two methods converge to comparable values.
We determine the radius, $R_{S2N}$, at which the signal-to-noise ratio is higher than 2,
and select the objects in the sample that satisfy the criterion $R_{S2N}/R_{200} > 0.7$. 
For the eleven selected objects, we model by a power-law function the behaviour of $S_b(r)$ 
to estimate the slope at several characteristic radii expressed as a fraction of $R_{200}$.
}
{We measure a consistent steepening of the $S_b(r)$ profile moving outward
from $0.4R_{200}$, where an average slope of $-3.6$ ($\sigma=0.8$) is estimated.
At $R_{200}$, we evaluate a slope of $-4.3$ ($\sigma=0.9$) that implies a slope
in the gas density profile of $\approx -2.6$ and a predicted mean value of the
surface brightness in the $0.5-2$ keV band of $2 \times 10^{-12}$ 
erg s$^{-1}$ cm$^{-2}$ deg$^{-2}$. 
}
{Combined with estimates of the outer slope of the gas temperature 
profile and expectations about the dark matter distribution, 
these measurements lie well within the physically allowed regions,
allowing us to describe properly how X-ray luminous
clusters behave out to the virial radius.
}


\maketitle


\section{Introduction}
Galaxy clusters form by the hierarchical accretion of cosmic matter.
The end products of this process are virialized structures that, in the X-ray band,
exhibit similar radial profiles of surface brightness (e.g. Vikhlinin et al. 1999, Neumann 2005), 
plasma temperature (e.g. Allen et al. 2001, Vikhlinin et al. 2005)
and dark matter distribution (e.g. Pointecouteau, Arnaud \& Pratt 2005).
These measurements have been improved due to the arcsec resolution and 
large collecting area of the X-ray satellites, such as
\chandra\ and \xmm, but still remain difficult because of the 
high signal-to-noise ratio required.
On the other hand, the X-ray surface brightness is a far easier quantity to observe
and define, which is rich in physical information being proportional
to the emission measure of the emitting source.
Recent work focused on a few local bright objects with {\it ROSAT} PSPC observations, 
which have low instrumental background and large field of view, to recover and
characterise the X-ray surface brightness profile over a significant fraction of the virial radius
(Vikhlinin et al. 1999, Neumann 2005).

In this work, we study the X-ray emission of a sample of
galaxy clusters in the redshift range $0.3-1.3$ observed with the arcsec
resolution of \chandra\ (see Balestra et al. 2007),
with the purpose of mapping, even at high redshift, a region out to $R_{200}$
and place reasonable constraints on the slope
of the gas density in the outskirts of hot ($kT > 3$ keV) galaxy clusters.

We assume a Hubble constant of 70 $h_{70}$ km s$^{-1}$ Mpc$^{-1}$ in a flat universe
with $\Omega_{\rm m}$ equal to 0.3.
All quoted errors are presented as the 1 sigma confidence level, unless otherwise stated.

\section{Preparation of the dataset}

\begin{figure*}
\hbox{
 \epsfig{figure=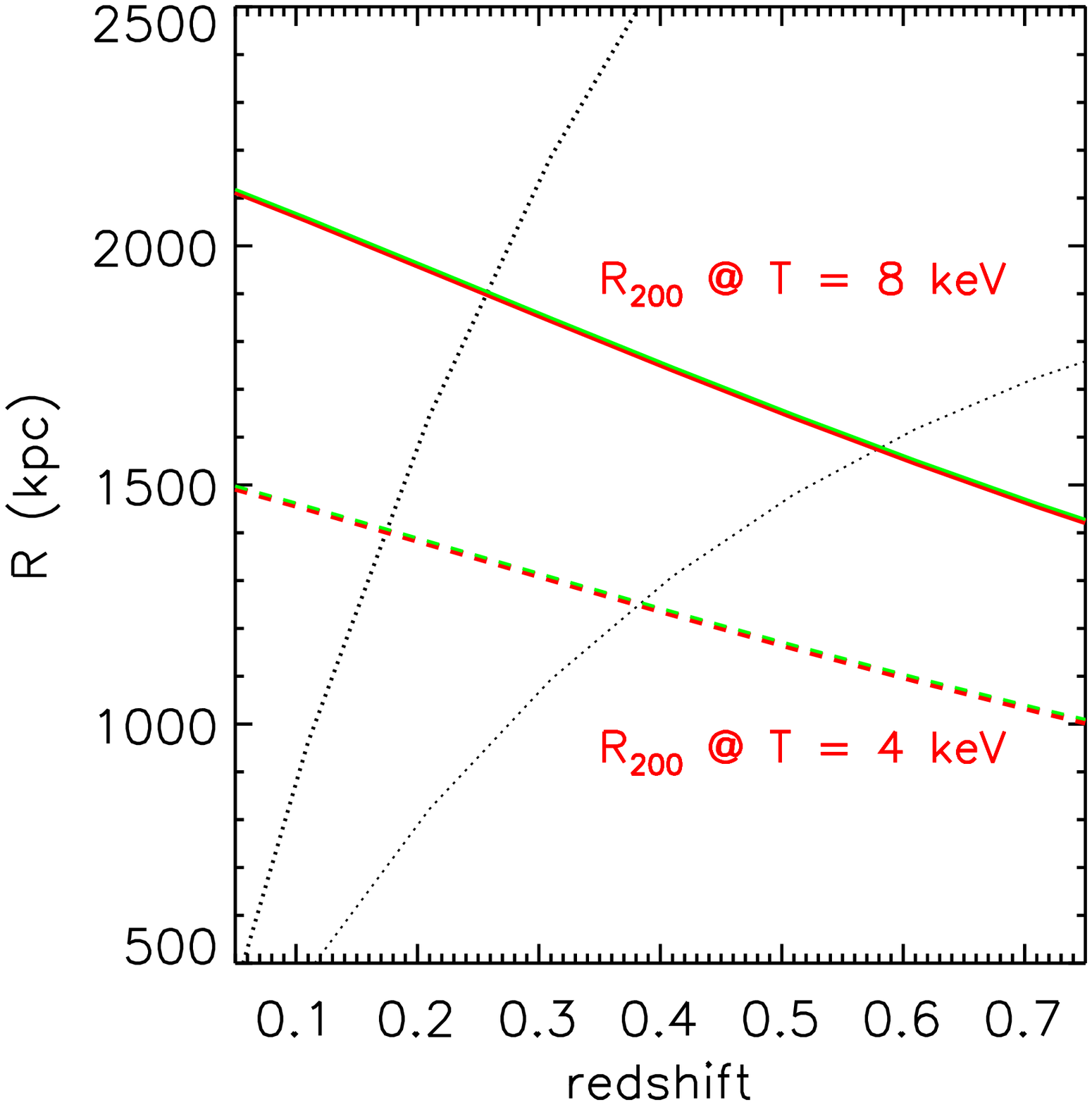,width=0.33\textwidth}
 \epsfig{figure=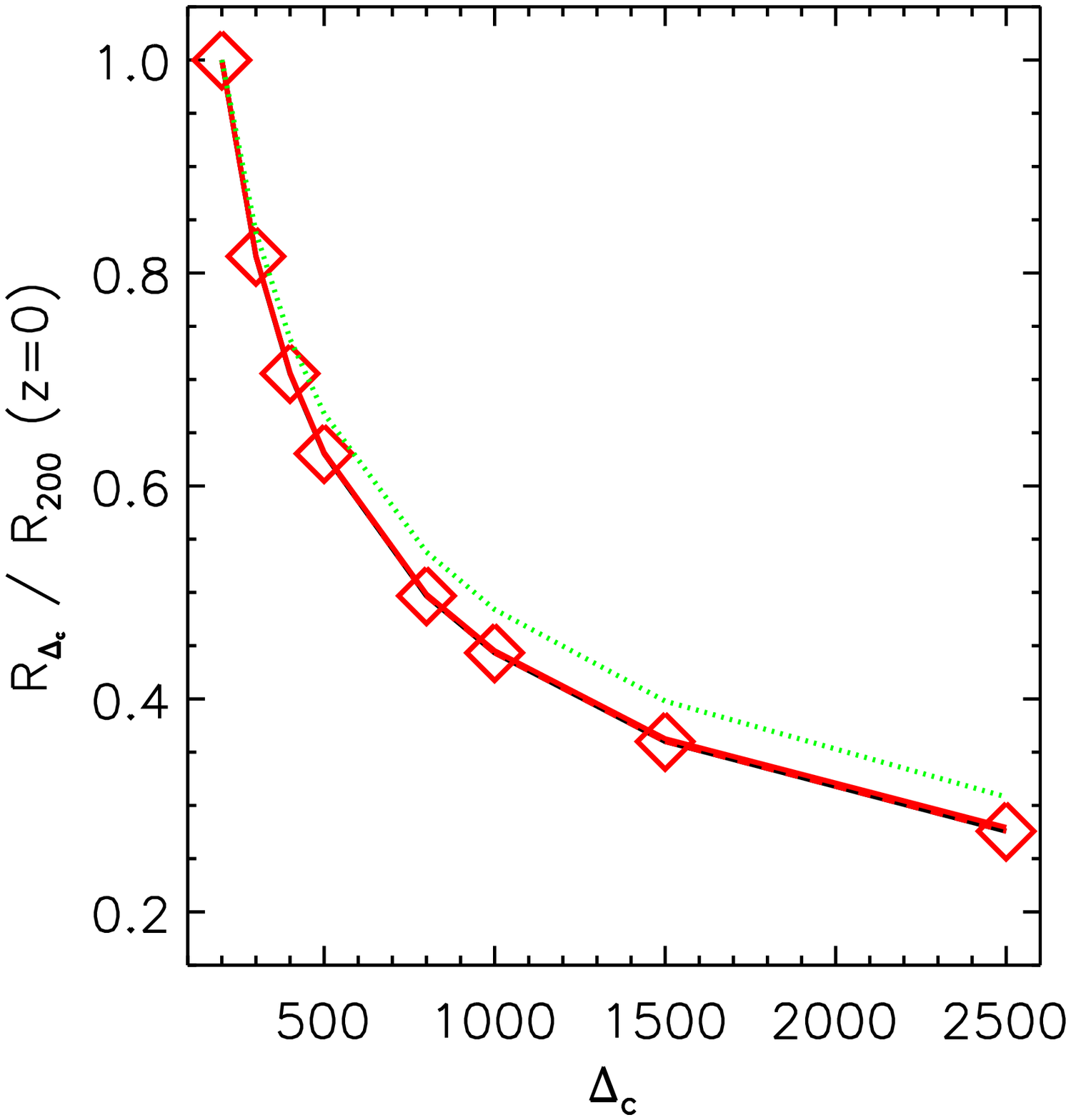,width=0.33\textwidth}
 \epsfig{figure=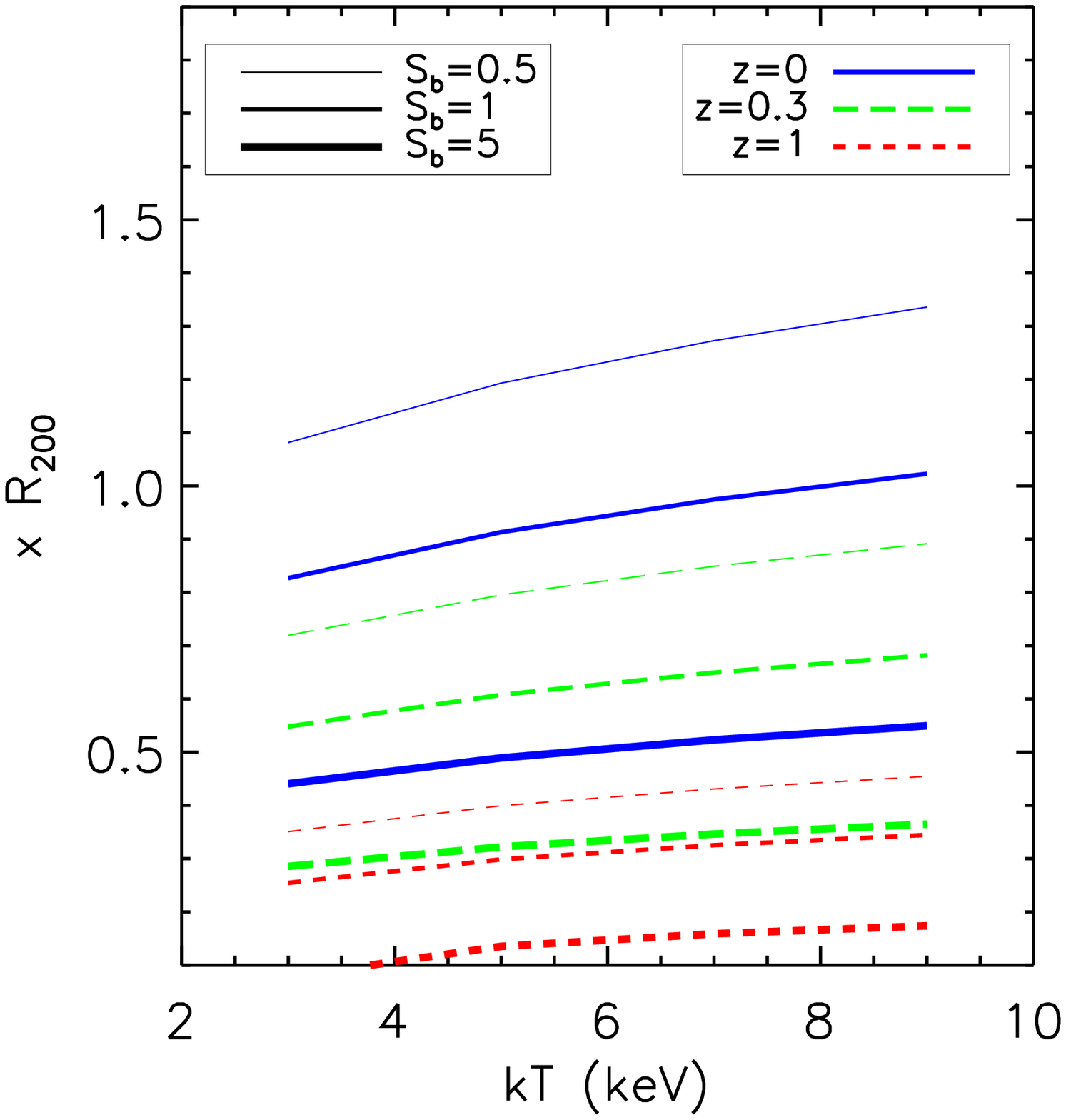,width=0.33\textwidth}
}
\caption{
({\bf Left}) Expected $R_{200}$ for an object with 4 ({\it dashed line}) 
and 8 keV ({\it solid line}) and radius corresponding to 4 and 8 ({\it thickest line}) 
arcmin for the assumed $\Lambda$CDM cosmology as function of the observed redshift.
Two estimates, almost indistinguishable, of $R_{200}$ are provided,
one obtained with a $\beta-$model with
core radius of 0.1 Mpc and $\beta=0.6$ ({\it thickest lines}), 
the other from the best-fit relation in Arnaud et al. (2005): 
$h_z R_{200} = 1714 (\pm 30) (T/5 {\rm keV})^{0.50\pm0.05}$. 
({\bf Center}) Fraction of $R_{200}$ mapped within a given overdensity
$\Delta_{\rm c}$ at redshift $0$ for a cluster with a gas temperature of
4 ({\it diamonds} and {\it dashed line}) and 8 keV ({\it solid line}).
The region at $\rho/\rho_{\rm c} = 500$ is enclosed within
a sphere with radius $\approx 0.65 \times R_{200}$.
The dotted line indicates the mapped fraction of $R_{200}$ assuming 
a Navarro-Frenk-White profile (e.g. Navarro et al. 2004) with concentration 
of 6, typical for the massive objects under consideration. 
({\bf Right}) Regions mapped (as fraction of $R_{200}$) for fixed surface 
brightness at different redshifts and gas temperatures, $S_b \sim
T^{0.5} (1+z)^{-4}$.
As reference, a value of $S_b =1$ at $R_{200}$, $z=0$, $kT=8$ keV and a
$\beta-$model with core radius of 0.1 Mpc and $\beta=0.6$ are adopted.
} \label{fig:r200}
\end{figure*}

\begin{table*}
\caption{Sample of galaxy clusters, which satisfies the criterion
$R_{S2N}/R_{200} > 0.7$. The columns show: the name of the cluster,
the observation IDs, in which ACIS configuration has been observed,
the exposure time, the adopted X-ray center,
the Galactic absorption from Dickey \& Lockman (1990) in correspondence
of the X-ray center, the redshift, the gas temperature, the mean
local background, $B$, in counts observed in the $0.5-5$ keV band, 
and the average distance $D_B$ of the regions used
to estimate the local background from the X-ray center.
Note: MACSJ0744.9+3927 is the merger of three exposures with 
IDs 3197, 3585, and 6111.
}
\begin{tabular}{l@{\hspace{.5em}} c@{\hspace{.5em}} c@{\hspace{.5em}} c@{\hspace{1.0em}} r@{\hspace{.7em}} r@{\hspace{.7em}} c@{\hspace{.4em}} c@{\hspace{.6em}} c@{\hspace{.7em}} c@{\hspace{.7em}} c@{\hspace{.7em}} }
\hline \\
cluster & Obs ID & ACIS & $t_{\rm exp}$ & RA & Dec & $n_H$ & $z$ & $kT$ & $B (\times 10^{-3})$ & $D_B$ \\
 & & & ks & h m s & degree & $10^{20}$ cm$^{-3}$ & & keV & cts/s/arcmin$^2$ & $R_{200}$ \\ 
 & & & \\ 
\hline \\
MS0015.9+1609 & 520      & I & 67.4 & 0 18 33.8 & +16 26 12 & 4.1 & 0.541 & $9.59\pm0.43$ & $1.61\pm0.03$ & $1.35$ \\
MACSJ2228.5+2036 & 3285     & I & 19.9 & 22 28 33.9 & +20 37 15 & 4.6 & 0.412 & $8.25\pm0.59$ & $2.00\pm0.07$ & $0.92$ \\
MACS0744.9+3927 & Merged   & I & 89.0 & 7 44 52.8 & +39 27 26 & 5.7 & 0.686 & $9.58\pm0.67$ & $1.67\pm0.04$ & $1.64$ \\
MACSJ0417.5-1154 & 3270     & I & 11.8 & 4 17 34.6 & -11 54 33 & 3.9 & 0.440 & $10.84\pm0.98$ & $2.33\pm0.10$ & $0.84$ \\
RXJ1701.3+6414 & 547      & I & 49.5 & 17 01 23.8 & +64 14 11 & 2.6 & 0.453 & $4.36\pm0.28$ & $2.25\pm0.05$ & $1.18$ \\
MACSJ1720.2+3536 & 3280     & I & 20.8 & 17 20 16.8 & +35 36 26 & 3.4 & 0.391 & $6.46\pm0.33$ & $1.87\pm0.09$ & $0.91$ \\
RXJ1416.4+4446 & 541      & I & 28.5 & 14 16 27.8 & +44 46 45 & 1.2 & 0.400 & $3.43\pm0.20$ & $2.22\pm0.06$ & $1.38$ \\
MACSJ2129.4-0741 & 3199     & I & 17.6 & 21 29 26.6 & -7 41 29 & 4.8 & 0.570 & $9.17\pm0.90$ & $1.88\pm0.09$ & $1.12$ \\
MACSJ1621.3+3810 & 3254     & I & 9.8 & 16 21 24.9 & +38 10 08 & 1.1 & 0.465 & $6.62\pm0.74$ & $1.58\pm0.10$ & $1.21$ \\
MS0451.6-0305 & 902      & S & 42.6 & 4 54 11.3 & -3 00 56 & 5.0 & 0.540 & $9.11\pm0.45$ & $3.98\pm0.07$ & $0.76$ \\
MACSJ1206.2-0847 & 3277     & I & 23.4 & 12 06 12.1 & -8 48 02 & 3.7 & 0.440 & $11.98\pm0.85$ & $2.07\pm0.06$ & $0.78$ \\
\hline \\ 
\end{tabular}

\label{tab:obj}
\end{table*}

To study the outskirts of X-ray emitting galaxy clusters, possibly out to the 
virial region, we require bright objects located at an appropriate redshift 
to allow the \chandra\ field-of-view to encompass the interesting area.
As shown in Fig.~\ref{fig:r200} (panel on the left) for a typical massive
object, we should expect to resolve the region within an
overdensity of 200, estimated with respect to the critical density at the
cluster redshift, using instruments with a field-of-view of radius
larger than 8 arcmin above redshift $0.2$.   
In the same figure, we plot the expected dependence of the radius
(as a function of $R_{200}$) on both the cluster overdensity (central panel)
and the gas temperature and redshift for a given surface brightness (panel
on the right). These plots indicate that any overdensity is mapped 
within a fixed fraction of $R_{200}$ regardless of the cluster temperature
and/or mass concentration
(e.g. $\Delta_{\rm c} = 500$ is reached at about $0.65 R_{200}$ for any
$T$ of interest, and with variations of 2 per cent with 
a concentration parameter in the range $4-8$ for a given 
Navarro-Frenk-White mass profile; e.g. Navarro et al. 2004),
and the surface brightness that a galaxy cluster is expected 
to emit for a given $T$ and $z$ with respect to the value measured
at $z=0$ in an object with $T=5$ keV (e.g. the same $S_{\rm b}$ value
is expected in a 10 keV system at $z\sim0.3$).

In the present work, we consider a sample of hot ($T_{\rm gas} > 3$ keV), 
high-redshift ($0.3 < z < 1.3$) galaxy clusters described in Balestra et al. (2007).
We refer to that work for details of the data reduction. We recall
that the spectral analysis was performed by extracting
the spectrum from a circular region of radius $\sim 0.15-0.3 R_{200}$ 
defined in order to maximize the signal-to-noise ratio in each cluster.
These regions contain the core emission, which has not been removed
due to the low count statistics for the high redshift systems under
consideration. On the other hand, results from observations and
numerical simulations (e.g. Santos et al. 2008, Ettori \& Brighenti 2008)
indicate a lower incidence of cooling cores in clusters at higher redshift,
suggesting that our overall estimates of $T_{\rm gas}$ should not be significantly 
biased towards low values. Moreover, the measurements of the gas temperature
were used in our analysis only to infer a physical radius, which depends on the square
root of $T_{\rm gas}$. Therefore, any error affecting our temperature estimates propagates
only half of its value as a  relative error on the physical radius of interest.

We prepare the exposure-corrected images in the energy band $0.5-5$ keV 
as described in Tozzi et al. (2003) and Ettori et al. (2004).
The exposure maps for each cluster were computed by combining different 
instrumental maps at given energy, using weights defined from a thermal spectrum 
with the best-fit parameters obtained from our spectral analysis.
The variations in the exposure map were not expected to be significant
in the adopted energy band 
\footnote{See, for example, Fig.~2a in http://cxc.harvard.edu/ciao3.4/download/doc/expmap\_intro.ps}.
We masked any detectable point sources with circular regions of radii 
large enough to include all the emission and, in any case, larger than 2 
arcsec.
We preferred to remove the point sources manually to avoid any possible 
confusion with detections by automatic algorithms (i.e. wavedetect) 
in proximity or within the extended emission of the clusters.

The center of the cluster was defined to correspond to the centroid of the raw image 
estimated in a box of width $\sim10-20$ arcsec around the maximum of the 
image itself that had been smoothed with a moving average
in a square box of width equal to $3-5$ arcsec.

The surface brightness profiles were extracted by requiring a fixed number of a
minimum 50 counts per bin in the inner 10 radial bins and then collecting the 
counts within annulii for which the outer radius was increased at each step 
by a factor of 1.2.

A local background, $B$, was defined for each exposure by considering 
a region far from the X-ray center that covered a significant portion
of the exposed CCD with negligible cluster emission.
The estimated average value of $B$, its relative error, and the mean distance 
from the X-ray center of the region analysed to determine the background
are quoted in Table~\ref{tab:obj}.

We define the ``signal-to-noise'' ratio, $S2N$, to be the ratio of the observed surface
brightness value in each radial bin, $S_b(r)$, after subtraction of the estimated background, 
$B$, to the Poissonian error in the evaluated surface brightness, 
$\epsilon_b(r)$, summed in quadrature with the error in the background, $\epsilon_B$:
$S2N(r) = \left[ S_b(r) - B \right] / \sqrt{\epsilon_b(r)^2 +\epsilon_B^2}$.
The outer radius at which the signal-to-noise ratio remained above $2$ was defined
to be the limit of the extension of the detectable X-ray emission, $R_{S2N}$.   
  
\begin{figure*}
 \epsfig{figure=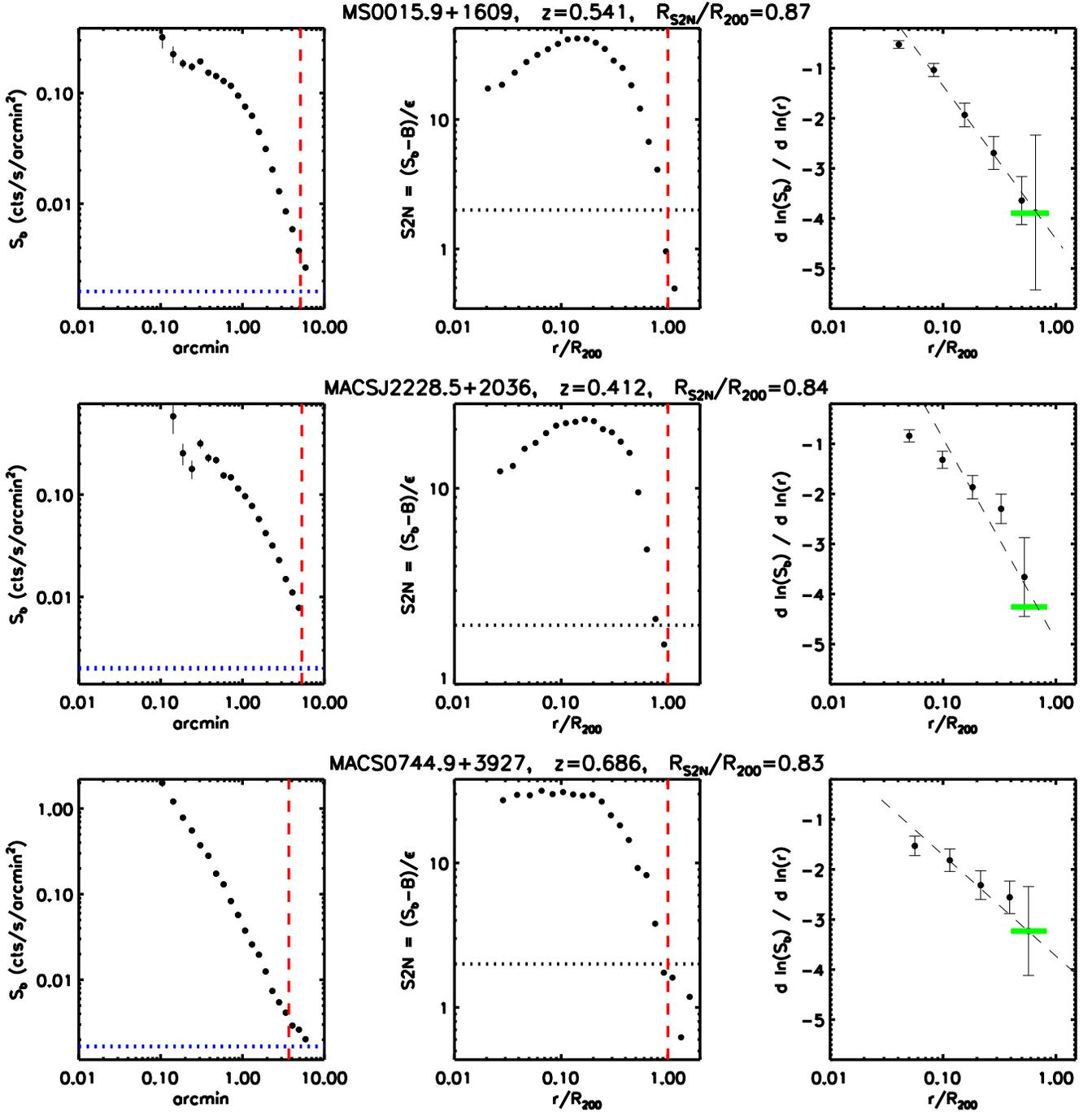,width=\textwidth}
 \caption{In the first column, we plot the surface brightness profiles
of the objects under examination with the fitted background
({\it horizontal dotted line}) and the radius $R_{200}$
({\it vertical dashed line}).
The panels in the second column show the signal-to-noise profiles, evaluated
as $S2N = (S_b - B)/ \epsilon$, where the error $\epsilon$ is the sum
in quadrature of the Poissonian error in the radial counts and the
uncertainties in the fitted background, $B$.
They are plotted as a function of $r / R_{200}$.
In the third column, we plot the best-fit values of the slope of the
surface brightness profiles as a function of $r / R_{200}$. These
values are estimated over 6 radial bins.
The thick horizontal solid line indicates the slope evaluated between
$0.4 \times R_{200}$ and $R_{S2N}$ with a minimum of 3 radial bins.
The dashed line indicates the best-fit of $d \ln (S_b) / d \ln (r/R_{200})$
with the functional form $s_0 +s_1 \ln (r/R_{200})$ over the radial range
$0.1 \times R_{200} - R_{S2N}$, with the best-fit
parameters quoted in Table~\ref{tab:fit}.
} \label{fig:s2n}
\end{figure*}

We note that the local background is about a factor of 2 higher
in the ACIS-S field than in ACIS-I.
We observe no significant correlation of $B$ with the Galactic absorption
($<6 \times 10^{20}$ cm$^{-2}$ in the selected objects; a random deviation
from the null value of no-correlation is expected with probability of 
$0.19$), indicating that we were dominated by cosmic and instrumental background.
A negligible (with probability of $0.69$) correlation
is also noticed between $B$ and the mean distance $D_B$ (see Table~\ref{tab:obj})
of the region from which the local background is extracted.
A slightly positive trend (significance of $1.6 \times 10^{-4}$)
was instead present between the estimated
uncertainty $\epsilon_B$ and the exposure time of the observation, as expected.

To scale the radial quantity with respect to a physical radius,
we estimated $R_{200}$ by using the single temperature measurement
(see Table~\ref{tab:obj}) and the best-fit description of $S_b(r)$ with a $\beta-$model
(the best-fit parameters are shown in Table~\ref{tab:s2n}, obtained
after fitting the background-subtracted profile over all positive values
above 40 kpc) following the equation in Ettori (2000):
\begin{equation}
R_{200} = r_{\rm c} \times \left[ \left( \frac{195.45 \, \beta \, \gamma \,
T_{\rm gas}}{\Delta_{\rm c} \, E_z^2 \, h_{70}^2 \, r_{\rm c}^2} \right)^a -1 \right]^{0.5},
\label{eqn:r200}
\end{equation}
where $E_z = \left[\Omega_{\rm m} (1+z)^3 +1 -\Omega_{\rm m} \right]^{0.5}$,
$\Delta_{\rm c}=200$, $a = \left[ 1.5 \beta (\gamma-1) +1 \right]^{-1}$, and
the polytropic index $\gamma$ is fixed to be equal to 1.

\begin{table*}
\caption{Best-fit results of the X-ray surface brightness profiles
modelled with a $\beta-$model.
$\overline{R_{200}}$ is estimated from the scaling laws in Arnaud et al. (2005);
$R_{200}$ from Eq.~\ref{eqn:r200}; $R_{S2N}$ and $r_{\rm c}$ are quoted
in unit of $R_{200}$.
}
\begin{tabular}{l@{\hspace{.7em}} c@{\hspace{.7em}} c@{\hspace{.7em}} c@{\hspace{.7em}} c@{\hspace{.7em}} c@{\hspace{.7em}} }
\hline \\
cluster & $\overline{R_{200}}$ & $R_{200}$ & $R_{S2N}$ & $r_{\rm c}$ & $\beta$ \\
 & kpc & kpc & $R_{200}$ & $R_{200}$ & \\ 
 & & & \\ 
\hline \\
MS0015.9+1609 & $1770$ & $1948\pm43$ & $0.871$ & $0.137\pm0.005$ & $0.742\pm0.011$ \\
MACSJ2228.5+2036 & $1771$ & $1743\pm62$ & $0.837$ & $0.093\pm0.007$ & $0.587\pm0.013$ \\
MACS0744.9+3927 & $1622$ & $1566\pm56$ & $0.832$ & $0.045\pm0.003$ & $0.561\pm0.005$ \\
MACSJ0417.5-1154 & $1998$ & $1936\pm91$ & $0.783$ & $0.068\pm0.006$ & $0.567\pm0.010$ \\
RXJ1701.3+6414 & $1258$ & $1204\pm44$ & $0.731$ & $0.087\pm0.013$ & $0.555\pm0.020$ \\
MACSJ1720.2+3536 & $1587$ & $1617\pm42$ & $0.726$ & $0.056\pm0.004$ & $0.626\pm0.012$ \\
RXJ1416.4+4446 & $1150$ & $1133\pm38$ & $0.722$ & $0.080\pm0.010$ & $0.587\pm0.020$ \\
MACSJ2129.4-0741 & $1701$ & $1684\pm83$ & $0.712$ & $0.069\pm0.006$ & $0.592\pm0.014$ \\
MACSJ1621.3+3810 & $1538$ & $1516\pm88$ & $0.711$ & $0.048\pm0.007$ & $0.585\pm0.019$ \\
MS0451.6-0305 & $1727$ & $1995\pm52$ & $0.705$ & $0.118\pm0.005$ & $0.814\pm0.015$ \\
MACSJ1206.2-0847 & $2100$ & $2158\pm77$ & $0.703$ & $0.067\pm0.004$ & $0.638\pm0.009$ \\
\hline \\ 
\end{tabular}

\label{tab:s2n}
\end{table*}

For comparison, we also estimated $R_{200}$ by adopting
the scaling relation with $T_{\rm gas}$ obtained
for massive clusters in Arnaud et al. (2005; similar
results were presented in Vikhlinin et al. 2005):
$\overline{R_{200}} = 1714 (\pm 30) \times (T_{\rm gas}/5
{\rm keV})^{0.50\pm0.05} h_{70}^{-1} E_z^{-1}$ kpc.
The value of $\overline{R_{200}}$ was estimated and quoted
in Table~\ref{tab:s2n}.
No appreciable differences were present between the different estimates of $R_{200}$,
such that the mean and standard deviation of the ratio of
the two values were $1.02$ and $0.06$, respectively.

\section{The surface brightness at $r > R_{500}$}

We investigated the X-ray surface-brightness profiles
of massive clusters at $r > R_{500} \approx 0.7 R_{200}$ (see Fig.~\ref{fig:r200}),
selecting the 11 objects with $R_{S2N}/R_{200} > 0.7$.
The properties of these clusters are shown in Tables~\ref{tab:obj} and \ref{tab:s2n}.
Examples of the analysed dataset are shown in Fig.~\ref{fig:s2n}.

We performed a linear least-squares fit between the logarithmic values of the 
radial bins and the background-subtracted X-ray surface brightness, 
taking into account the data errors in both coordinates
with the routine {\it FITEXY} (Press et al. 1992; as implemented in IDL). 

\begin{figure}
 \epsfig{figure=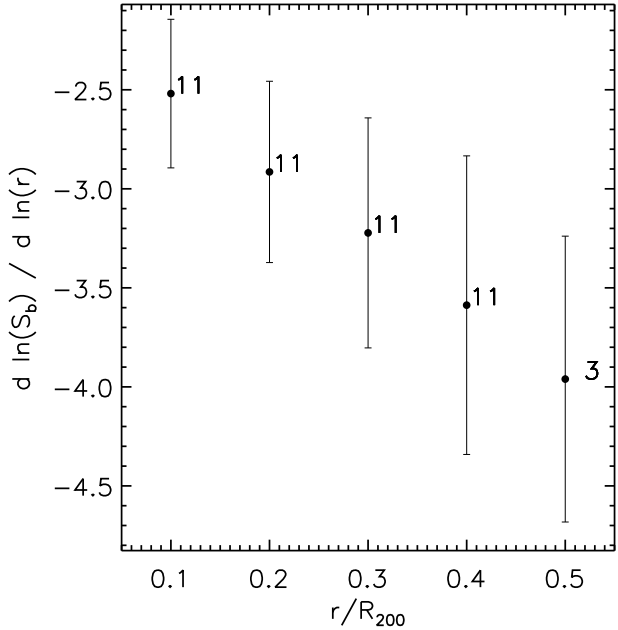,width=0.5\textwidth}
\caption{Error-weighted mean and standard deviation of the distribution
of the values of the slope of the surface brightness in the radial range
$r - R_{S2N}$ plotted as a function of the inner radius $r$.
The number of objects with at least 3 points
considered for this estimate is indicated.
} \label{fig:slope_r}
\end{figure}

The distribution of the error-weighted mean slopes above a fixed fraction
of $R_{200}$ and within $R_{S2N}$ is shown in Fig.~\ref{fig:slope_r} and
quoted in Table~\ref{tab:fit}.
Overall, the error-weighted mean slope is $-2.91$ (with a standard deviation
in the distribution of $0.46$) at $r > 0.2 R_{200}$ and $-3.59 (0.75)$
at $r > 0.4 R_{200}$.
For the only 3 objects for which a fit between $0.5 R_{200}$ and $R_{S2N}$
was possible, we measured a further steepening of the profiles, with a
mean slope of $-4.43$ and a standard deviation of $0.83$.

We also fitted linearly the derivative (from numerical differentiation 
using three-point Lagrangian interpolation as described in Hildebrand 1987 and 
implemented in the IDL function {\it DERIV}) of the logarithm $S_b(r)$ 
over the radial range $0.1 R_{200} - R_{S2N}$, excluding in this way the
influence of the core emission
(see values $s_0$ and $s_1$ in Table~\ref{tab:fit}).
The average (and standard deviation $\sigma$) values of the extrapolated slopes are 
then $-3.15 (0.46)$, $-3.86 (0.70)$, and $-4.31 (0.87)$ 
at $0.4 R_{200}$, $0.7 R_{200}$ and $R_{200}$, respectively.

These values are comparable to these obtained in previous analyses.
Vikhlinin et al. (1999) found that a $\beta-$model with $\beta=0.65-0.85$
described the surface brightness profiles in the range $0.3-1 R_{180}$ of
39 massive local galaxy clusters observed with {\it ROSAT} PSPC.
These values translate in a range of the estimate of the power-law slope 
($-2.9/-4.1$, from $S_b(r) \approx r^{2 (0.5 -3\beta)} = r^{1 - 6 \beta}$) that includes
our estimates.

Neumann (2005) found that the stacked profiles of a few massive nearby systems
located in regions of low ($<6 \times 10^{20}$ cm$^{-3}$) Galactic absorption
observed by {\it ROSAT} PSPC still provided values of $\beta$ around $0.8$ at
$R_{200}$, with a power-law slope that increased from $-3$, when the fit was
done over the radial range $[0.1, 1] R_{200}$, to $-5.7^{+1.5}_{-1.2}$ over
$[0.7, 1.2] R_{200}$.

These observational results were supported by hydrodynamical simulations
of X-ray emitting galaxy clusters
performed with the Tree+SPH code GADGET-2 from Roncarelli et al. (2006).
In the most massive systems, they measured a steepening of $S_b(r)$, independently
from the physics adopted in treating the baryonic mass, with a slope of
$-4, -4.5,$ and $-5.2$ when estimated for the radial range $0.3-1.2 R_{200}$,
$0.7-1.2 R_{200}$, and $1.2-2.7 R_{200}$, respectively.
In particular, we note the good agreement between the slope of the simulated
surface brightness profile of the representative massive cluster
in the radial bin $0.7-1.2 R_{200}$ (see values of $b_A$ in Table~4 of
Roncarelli et al. (2006), ranging between $-4.29$ and $-4.54$)
and our mean extrapolated value at $R_{200}$ of $-4.31$.

In Table~\ref{tab:fit}, we quote the estimated surface brightness value
at $R_{200}$ in CGS units. They were obtained by extrapolating the
measured $S_{\rm b}$ at $\sim 0.7 R_{200}$ out to $R_{200}$ by using
the best-fit values quoted in the same Table. The conversion from the 
count rate to the flux was completed by adopting an absorbed thermal component 
at the cluster redshift with an assumed metallicity and temperature equal to 
(i) the ones measured and (ii) one third of the measured values.
We predict an average surface brightness of about $2 \times 10^{-12}$
erg s$^{-1}$ cm$^{-2}$ deg$^{-2}$ in the $0.5-2$ keV band, corresponding 
to about half of the value associated with the total
cosmic background observed (e.g. Hickox \& Markevitch 2006) and consistent
with the estimates measured in the hydrodynamical simulations of massive
galaxy clusters (Roncarelli et al. 2006).
Converting this surface brightness to counts in the $0.5-5$ keV band,
we predict that at $R_{200}$ the X-ray emission from a cluster
is responsible of $\sim 5$ per cent of the total counts 
observable with \chandra.

\begin{table*}
\caption{Best-fit values of the slope of the surface brightness 
profile in the radial range $[0.2 \times R_{200}, R_{S2N}]$ and 
$[0.4 \times R_{200}, R_{S2N}]$. 
The best-fit results of the first derivative of $S_b(r)$
over the entire radial range are indicated as 
$d \ln (S_b) / d \ln (r/R_{200}) = s_0 +s_1 \ln (r/R_{200})$
and shown in Fig.~\ref{fig:s2n}. 
The predicted slopes from the best-fit results at $0.4, 0.7$, and 
$1 \times R_{200}$ are quoted.
In the last two columns, we present the expected surface brightness 
in the $0.5-2$ keV band and the fraction $fc_{200}$ of the clusters counts with 
respect to the total count rate measured in the $0.5-5$ keV at $R_{200}$.
}
\begin{tabular}{l@{\hspace{.7em}} c@{\hspace{.7em}} c@{\hspace{.7em}} c@{\hspace{.7em}} c@{\hspace{.7em}}
   c@{\hspace{.7em}} c@{\hspace{.7em}} c@{\hspace{.7em}} c@{\hspace{.7em}} c@{\hspace{.7em}} c@{\hspace{.7em}}}
\hline \\ 
cluster & & \multicolumn{2}{c}{slope} & & $(s_0, s_1)$ & $0.4 R_{200}$ & $0.7 R_{200}$ & $R_{200}$ & $S_{\rm b} (R_{200})$ & $fc_{200}$  \\ 
  & & $0.2 R_{200} - R_{S2N}$ & $0.4 R_{200} - R_{S2N}$ & & & & & & $10^{-12}$ erg/s/cm$^2$/deg$^2$ \\ 
 & & & &\\ 
\hline \\ 
MS0015.9+1609 & & $-3.30\pm0.28$ & $-3.89\pm0.96$ & & $(-4.41, -1.33)$ & $-3.19$ & $-3.93$ & $-4.41$ & $(1.7,1.7)$ & $0.051$ \\
MACSJ2228.5+2036 & & $-2.74\pm0.29$ & $-4.26\pm1.23$ & & $(-4.92, -1.75)$ & $-3.32$ & $-4.30$ & $-4.92$ & $(2.6,2.7)$ & $0.057$ \\
MACS0744.9+3927 & & $-2.73\pm0.31$ & $-3.23\pm0.89$ & & $(-3.73, -0.88)$ & $-2.92$ & $-3.41$ & $-3.73$ & $(2.6,2.7)$ & $0.074$ \\
MACSJ0417.5-1154 & & $-3.38\pm0.42$ & $-4.53\pm1.27$ & & $(-5.60, -2.02)$ & $-3.75$ & $-4.88$ & $-5.60$ & $(1.6,1.6)$ & $0.034$ \\
RXJ1701.3+6414 & & $-2.39\pm0.27$ & $-2.59\pm1.17$ & & $(-4.24, -1.60)$ & $-2.78$ & $-3.67$ & $-4.24$ & $(2.4,2.6)$ & $0.051$ \\
MACSJ1720.2+3536 & & $-3.19\pm0.36$ & $-2.40\pm1.32$ & & $(-3.57, -0.77)$ & $-2.87$ & $-3.30$ & $-3.57$ & $(2.5,2.6)$ & $0.059$ \\
RXJ1416.4+4446 & & $-2.74\pm0.37$ & $-3.63\pm1.72$ & & $(-3.75, -0.94)$ & $-2.89$ & $-3.41$ & $-3.75$ & $(1.8,1.9)$ & $0.042$ \\
MACSJ2129.4-0741 & & $-2.68\pm0.33$ & $-4.16\pm1.74$ & & $(-3.92, -1.13)$ & $-2.88$ & $-3.51$ & $-3.92$ & $(2.4,2.5)$ & $0.060$ \\
MACSJ1621.3+3810 & & $-2.64\pm0.37$ & $-3.53\pm1.76$ & & $(-2.99, -0.45)$ & $-2.58$ & $-2.83$ & $-2.99$ & $(3.0,3.2)$ & $0.087$ \\
MS0451.6-0305 & & $-3.94\pm0.45$ & $-4.77\pm2.04$ & & $(-5.85, -1.91)$ & $-4.11$ & $-5.17$ & $-5.85$ & $(0.3,0.3)$ & $0.006$ \\
MACSJ1206.2-0847 & & $-3.24\pm0.37$ & $-3.67\pm1.64$ & & $(-4.43, -1.16)$ & $-3.37$ & $-4.02$ & $-4.43$ & $(1.4,1.5)$ & $0.032$ \\
 & & & &\\ 
 mean & & $-2.91$ & $-3.59$ & & $(-4.31,-1.27)$ & $-3.15$ & $-3.86$ & $-4.31$ & $(2.0,2.1)$ & $0.050$ \\
 $\sigma$ & & $0.46$ & $0.75$ & & $(0.87,0.50)$ & $0.46$ & $0.70$ & $0.87$ & $(0.8,0.8)$ & $0.022$ \\
\hline \\ 
\end{tabular}

\label{tab:fit}
\end{table*}

\section{Physical properties of the outer cluster regions}

We combine the measured slope of the 
surface brightness in the cluster outskirts with the predicted
slopes of the gas temperature and dark matter profiles.
We have demonstrated that the slope depends on the radius at which it
is measured. Thus, we focus hereafter on the properties at $R_{200}$,
by representing with power-laws the radial dependence of the gas
density, $n_{\rm gas}$, temperature, $T_{\rm gas}$, and
dark matter, $\rho_{\rm DM}$, profiles:
\begin{equation}
\left\{ \begin{array}{l}
n_{\rm gas} \propto r^a  \\
T_{\rm gas} \propto r^b  \\
\rho_{\rm DM} \propto r^c.
\end{array}
\right.
\label{eqn:slopes}
\end{equation}

The surface brightness, $S_b$, is the integral along the line of sight
of the square of the gas density and can be written, 
assuming spherical geometry, as 
\begin{equation}
S_b = \int n_{\rm gas}^2 \Lambda(T) dl \ \propto n_{\rm gas}^2 \Lambda(T) r
  \ \propto r^{2 a \ +1 \ +0.2 b},
\label{eqn:sb}
\end{equation}
where we consider the weak dependence ($T \propto r^{0.2b}$) on the gas temperature 
of the cooling function, $\Lambda(T)$, due to the limited energy band $\Delta E = [0.5,5]$ keV
in which the images were considered and the gas temperatures under consideration (e.g. Ettori 2000).

Hydrostatic polytropic gas can be described by the relation
$T_{\rm gas} \propto n_{\rm gas}^{\gamma-1}$, with a polytropic index
$\gamma$ that relates the ratio of specific heats in a adiabatic gas and
equals $5/3$ for a monoatomic ideal gas. If the intracluster
gas is not entirely adiabatic, the polytropic index can be used
as a fitting parameter to describe the relative behaviour of the
gas temperature and density profile.
In particular, if the thermal conduction is an efficient process, the gas
tends to be isothermal and $\gamma=1$, whereas the entropy per atom will be
constant if the thermal conduction is slow and the gas is well mixed.
Moreover, when $\gamma > 5/3$, the gas becomes convectively unstable and mixing
occurs within several sound crossing times, which are in general quite short
compared to the overall age of the cluster.    
Therefore, this ``polytropic condition'' requires that $1 \le \gamma \le 5/3$
and translates into the request that
\begin{equation}
0 \le \frac{b}{a} \le \frac{2}{3}.
\end{equation} 

Further constraints are provided by the integrated masses.
The gas mass estimate, $M_{\rm gas} \approx \int n_{\rm gas} dV \ \propto 
r^{a+3}$, requires that 
\begin{equation}
a + 3 > 0.
\end{equation}
The total gravitating mass can be expressed both as an integral
of the dark matter profile over the cluster volume 
\begin{equation}
M_{\rm grav} = \int \rho_{\rm DM} dV \ \propto r^{c+3},
\end{equation}
and with the assumption of hydrostatic equilibrium between dark matter 
and spherically symmetric intracluster gas 
\begin{equation}
M_{\rm hyd} \approx - r \ T \ \left(
\frac{d \ln n_{\rm gas}}{d \ln r} +
\frac{d \ln T_{\rm gas}}{d \ln r} \right) \propto - (a+b) \ r^{b+1}. 
\end{equation}
The request for a gravitational mass increasing radially implies that $c + 3 >0$ and
$a + b <0$. Moreover, by equating the two definitions of the gravitational mass, 
we find that the exponents of the radial dependence must to satisfy 
the condition $c+3=b+1$, which directly relates the slope of the dark matter
profile to that of the gas temperature profile.

\begin{figure}
\epsfig{figure=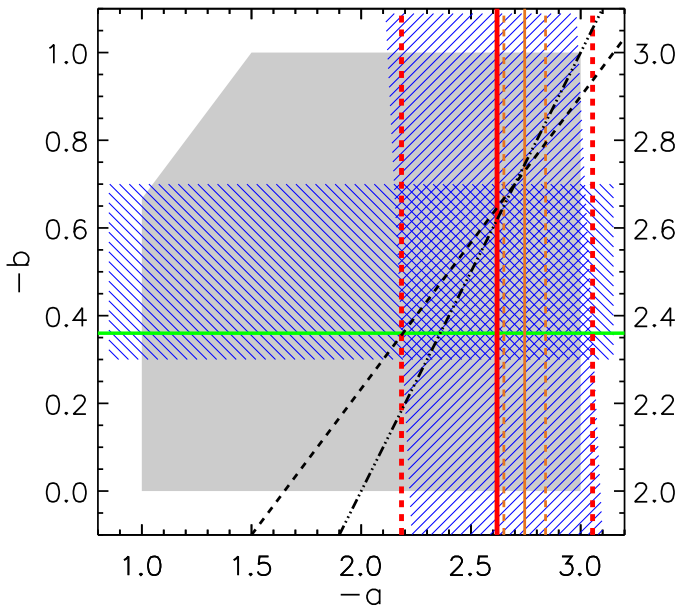,width=0.5\textwidth}
\caption{Constraints on the outer slopes of the gas density (index $a$), 
temperature (index $b$) and dark matter (index $c = b - 2$) profiles. 
The shaded area shows the physically allowed regions for
the index values as discussed in the text. 
The hatched regions indicate the $1 \sigma$ constraints from 
Navarro et al. (2004; horizontal regions) and from the present
work including the dependence upon the cooling function
(vertical area).
The horizontal green solid line shows the fiducial slope of the temperature 
profile from Vikhlinin et al. (2005).
The vertical solid (and dashed) lines indicate the limits on the slope 
of the gas density profile from hydrodynamical simulations (tighter
constraints in orange; Roncarelli et al. 2006) and from the present
observational analysis (in red).
The diagonal {\it dashed} line is the locus of the predicted
behaviour of the entropy profile, $b - 2a/3 = 1.1$.
The {\it dot-dot-dot-dashed} line satisfies the condition that 
the gas fraction is constant in the cluster outskirts.
} \label{fig:slope}
\end{figure}

Overall, these conditions can be summarized by the following
inequalities among the indices of the assumed power-laws:
\begin{equation}
\left\{ \begin{array}{l}
0 \le \frac{b}{a} \le \frac{2}{3} \\
a + 3 > 0  \\
c + 3 > 0  \\
a + b < 0  \\
c = b - 2 
\end{array}
\right.
\label{eqn:cond}
\end{equation}
We show in Fig.~\ref{fig:slope} the allowed regions in the $a-b-c$ plane.

It is worth noting that the permitted values shown in these plots 
also satisfy the requests
that the sound-crossing time, $t_{SC} \propto T_{\rm gas}^{-1/2} r
\approx r^{1 - b/2}$, the cooling time, $t_{cool} \propto 
T_{\rm gas}^{1/2} n_{\rm gas}^{-1} \approx r^{b/2 - a}$,
and the equipartition time for Coulomb collisions, $t_{eq} \propto 
T_{\rm gas}^{3/2} n_{\rm gas}^{-1} \approx r^{3b/2 - a}$,
increase with radius in the cluster outskirts.

Some observational constraints are available and can be located in these
graphs: (1) the mean value of the slope in the surface brightness at
$R_{200}$ is $-4.31 (\sigma=0.87)$, equal to $2 a \ + \ 1 \ + \ 0.2 b$ 
as inferred by Eq.~\ref{eqn:sb}, and implies that $a = -2.62 \pm 0.43$; 
(2) the dark matter profile behaves as $r^{-2.5 \pm 0.2}$ as discussed 
and shown in Fig.~3 of Navarro et al. (2004); 
(3) a polytropic index $\gamma \approx 1.24$ is measured in the outskirts of 
temperature profiles (e.g. Markevitch et al. 1998, De Grandi \& Molendi 2002
out to about $0.6 R_{180} \sim 0.63 R_{200}$;
a value of $1.42 \pm 0.03$ is estimated in one of the first spectral determination 
of the temperature profile at $R_{200}$, obtained for the cluster PKS0745-191 
with {\it Suzaku} in George et al. 2008).
These further constraints are also shown in Fig.~\ref{fig:slope}.

The expected slope of the dark matter profile, combined with
the observed slope in surface brightness, places constraints
on the predicted behaviour of the temperature profile (Eq.~\ref{eqn:cond})
and its polytropic index, $\gamma = 1 + b/a$. For the values quoted above,
we expect that $b \approx -0.5 \pm 0.2$ and $\gamma \approx 1.19 \pm 0.08$.
These values agree well with the constraints obtained from observed data-sets
(e.g. Markevitch et al. 1998, Vikhlinin et al. 2005, Leccardi \& Molendi 2008, 
Reiprich et al. 2008, George et al. 2008), hydrodynamical simulated objects
(e.g. Loken et al. 2002, Roncarelli et al. 2006) and analytical model of the ICM
(e.g. Ostriker, Bode \& Babul 2005).

Finally, the entropy profiles, $K(r) = T_{\rm gas}(r) / n_{\rm gas}^{2/3}$,
both from smooth and hierarchical accretion models 
(e.g. Tozzi \& Norman 2001, Voit 2005)
are predicted to increase with radius as $K \propto r^{1.1}$.
This shape of the entropy profile outside the core, but well within the
virial radius, is also observed in high quality cluster
exposures with the \xmm\ satellite (Pratt \& Arnaud 2003).
By using our power-law expressions, we find that 
$K(r) \propto r^{b  - 2 a/3}$, with a predicted value of the slope
of $1.25 \pm 0.35$ and $1.12$ for the simulated dark matter distribution
and the observed polytropic index, respectively.

\section{Summary and Conclusions}

We have selected 11 massive galaxy clusters in the redshift range $0.3< z <0.7$
observed with \chandra, which is convenient to survey the X-ray emission 
out to  a significant fraction of the virial radius and, in all the cases,
to radii beyond $R_{500} \approx 0.65 \times R_{200}$.

By fitting a single power-law function to the X-ray surface brightness profile $S_b(r)$ 
in different radial ranges, we have detected a consistent steepening in $S_b(r)$ 
moving outward, with a mean slope of $-2.9 \pm 0.5$ when the radial range 
$0.2R_{200}-R_{S2N}$ is considered, and $-3.6$ (r.m.s. $0.8$) when 
the lower limit of the radial range is $0.4R_{200}$.
The mean slope estimated from the linear fit to the derivative of $S_b(r)$
is $-3.2$, $-3.9$, and $-4.3$ at $0.4R_{200}$, $0.7R_{200}$, and $R_{200}$,
respectively. These values, corrected by the weak dependence of the cooling
function on gas temperature in the considered energy band $[0.5,5]$ keV, 
imply a slope in the gas density profile of $-2.04 \pm 0.23$ ($\beta \approx 0.68$), 
$-2.39 \pm 0.35$ ($\beta \approx 0.80$) and $-2.62 \pm 0.43$ ($\beta \approx 0.87$) 
at $0.4R_{200}$, $0.7R_{200}$ and $R_{200}$, respectively.

Combined with previous estimates of the outer slope of the gas temperature
profile (e.g. Vikhlinin et al. 2005, from an observational point of view;
Roncarelli et al. 2006, for hydrodynamically simulated clusters)
and the expectations for the dark matter distribution (e.g. Navarro et al. 2004),
and modelling of the gas density, temperature and dark matter profiles 
with single power laws ($n_{\rm gas} \propto r^a, T_{\rm gas} \propto r^b, 
\rho_{\rm DM} \propto r^c$), 
we have defined a physically-allowed region in the $a-b-c$ plane, with a 
predicted behaviour at $R_{200}$ described by power-law indices of
$a \approx -2.6, b \approx -0.5,$ and $c \approx -2.5$.

\section*{ACKNOWLEDGEMENTS}
We acknowledge the financial contribution from contracts ASI-INAF
I/023/05/0 and I/088/06/0.
Paolo Tozzi and Silvano Molendi are thanked for reading the manuscript 
and providing useful comments.
We thank the anonymous referee for very useful comments that improved
the presentation of the work.

\end{document}